\documentclass[%
 reprint,
 amsmath,amssymb,
 aps,
]{revtex4-2}

\usepackage{graphicx}% Include figure files
\usepackage{dcolumn}% Align table columns on decimal point
\usepackage{bm}% bold math
\usepackage{amsmath}
\usepackage{epsfig}
\begin{document}

\preprint{APS/123-QED}

\title{Spectroscopy and decay properties of bottomonium using \\ instanton induced potential from QCD vacuum}

%\author{Bhoomika Pandya}
%\email[]{bhumispandya@gmail.com}
%\affiliation{Department of Physics, Sardar Patel University,Vallabh Vidyanagar, INDIA}

%\author{Manan Shah}
%\email[]{mnshah09@gmail.com}
%\affiliation{P. D. Patel Institute of Applied Sciences, CHARUSAT, Changa, INDIA}

%\author{P C Vinodkumar}
%\email{p.c.vinodkumar@gmail.com}


\author{Bhoomika Pandya$^{1}$}
\email[]{bhumispandya@gmail.com}

\author{Manan Shah$^2$}
\email[]{mnshah09@gmail.com}

\author{P C Vinodkumar$^1$}
\email{p.c.vinodkumar@gmail.com}

\affiliation{$^1$Department of Physics, Sardar Patel University,Vallabh Vidyanagar, INDIA}
\affiliation{$^2$P. D. Patel Institute of Applied Sciences, CHARUSAT, Changa, INDIA}

\date{\today}% It is always \today, today,

\begin{abstract}
Ground state and radial excitation mass spectra of bottomonium states are computed using the Instanton Induced $q \bar{q}$ Potential obtained from Instanton Liquid Model for QCD vacuum. The Schrodinger equation is solved for two body problem using  variational method.  Hyperfine interactions are incorporated to remove the degeneracy between spin singlet and spin triplet states of $S$ wave masses. Spin- orbit and tensor interactions through one gluon exchange has been added to the spin average masses of P and D wave states.  The vector decay constants and pseudoscalar decay constants are also estimated using the Van Royan Weiskopff formula within non-relativistic limit. The di-leptonic, di-gamma, di-gluon decays of the bottomonium states are predicted without and with QCD corrections up to the lowest order using numerical values of the wave function at origin. Tri-gluon decays of respective $S$, $P$ and $D$ states are also studied. Other annihilation decay channels of bottomonium states into $\gamma gg$ , $\gamma \gamma \gamma$ , $q\bar{q} + g  $ are also computed.
The  radiative transitions of first order like  electric dipole transitions (E1) and magnetic dipole transitions (M1) are estimated. Our estimations of the mass spectra and decay properties for bottomonium found to be in excellent agreement with the average experimental values reported by particle data group.   
\end{abstract}

\maketitle

%\tableofcontents

\section{\label{sec:level1} Introduction}

Quarkonia are regarded as the simple and appropriate hadronic systems to explore the QCD aspects at the low energy regimes through its spectroscopy \cite{Eichten1978, A.Garmash2015, Eichten2008, Godfrey1985}. Since the discovery of $\Upsilon(1S)$, $\Upsilon(2S)$ and $\Upsilon(3S)$ at the Fermilab by E288 collaboration \cite{Herb1977, Innes1977}; many of its orbital excited states such as $\chi_{b0,1,2}  (1P)$ and $\chi_{b0,1,2}  (2P)$ etc. are also discovered subsequently \cite{Han1982, Eigen1982, Klopfenstein1983, Pauss1983}. Further, continuous progress has been achieved in the bottomnium spectroscopy by the discovery of $\Upsilon(4S)$, $\Upsilon(10860)$, $\Upsilon(11020)$ \cite{Besson1985, Lovelock1985}. BABAR Collaboration discovered the spin singlet low lying $\eta_{b}(1S)$ \cite{Aubert2008} and by the efforts of CLEO, Belle and BABAR its average mass is found to be 9399.0 $\pm$ 2.3  \cite{Bonvicini2010, Mizuk2012, Lees2011}. CLEO has given the successful observation of the $\eta_{b}(2S)$ with 9974.6 $\pm$ 2.3 $\pm$ 2.1 MeV \cite{Dobbs2012}but BELLE recorded the state having a mass of 9999.0 $\pm$ 3.5 after performing 17 times the same experiment but not for single time it founds signal nearer to 9974 MeV \cite{Sandilya2013}. $\Upsilon(1^3D_2)$ also well established \cite{Bonvicini2004, Sanchez2010}and low lying $h_{b}(1P)$ and $h_{b}(2P)$ are also detected by BABAR \cite{Lees20112}. Recently, LHCb collobaration found the mass of $\chi_{b 1}(3P)$ as $10515^{+2.2}_{-3.9}$ \cite{Aaij2014}. Two other charged states in the  bottom family, $Z_{b}(10610)$ and $Z_{b} (10650)$  are also being reported recently \cite{Bondar2012}. Their electric charge suggest that they are not conventional bottomonium \cite{Guo, Danilkin2012}, so we do not include them for the present work.

Looking to the advances in the experimental side, it is necessary to review earlier theoretical attempts to understand the quarkonia systems. Although there are many different approaches like Lattice QCD methods \cite{Hughes2015, Becirevi2015, Lewis2011, Baker2015}, NRQCD methods \cite{Brambilla2006, Brambilla2012, Pineda2013}, Light front quark model \cite{Choi2007, Ke, Ke2010, Ke2013}, various coupled channel quark models \cite{Ferretti2014, Ferretti20142, Lu2016}. Effective Lagrangian approach \cite{Fazio2011} etc., employed to study them, still there is no consensus and discrepancy persist in the predictions. For instance, authors \cite{Segovia2016} have used nonrelativistic constituent quark model wherein the lowest lying bottomonium states $\Upsilon(1S)$ and $\eta_{b}(1S)$ are about 50 MeV heigher than the PDG \cite{PDG2018} data but the same model give fine agreement for the higher excited state $\Upsilon(6S)$ while the relativized quark model \cite{Godfrey2015} provides very good discription of the low-lying states but higher excited $\Upsilon (6S)$ is overestimated by 100 MeV.

It is believed that most reliable description of the bound system is given by Bethe-Salpeter formalism. It is a relativistic quantum field theory but ambiguity is that the interaction kernel is  not solvable from QCD and it becomes very tough to extract hadronic information \cite{Lucha1991}. In EFTs, threshold region is still disappointing \cite{Brambilla2011}. Lattice regularized QCD beyond threshold and full understanding of bottomonium seems difficult \cite{Dudek2008, Meinel2009, Gray2005,  Burch2010, Meinel2010, Lewis2012, Wurtz2015,Dowdall2012} even though very recently they have predicted few radial excitations also. 

Currently, computational complexity of other theoretical approaches makes the potential model approach as the most reliable which is able to meet the expectations. Potential models provide quantitative and qualitative analysis of the quarkonia states. Potential models account for both the shorter and large distance behaviour of the $q$-$\bar{q}$ interaction. The most promising non-perturbative contibution is the "Quark Confinement" which can be incorporated by the ``Wilson loop``. Such non-perturbative effects will increase at the larger distances \cite{Eichten1975, Eichten1978}. And such attempts are key to the lattice QCD calculations. Apart from the Wilson loop, the contribution from Instantons \cite{Belavin1975} is also considered for the non-perterbative effects at the larger distance. Instantons, which are the large fluctuions of the gluon field and corresponds to the tunneling from one minimum of the energy to the neighbouring one. Similar fluctuations but having tunneling in the opposite side is called "anti-instantons".

In the present work we aim to find the effect of the instantons on the heaviest quarkonium system. As the form of the instanton potential contains the term which gives the non-peerturbative effect at the larger distance as well as coloumbian type behaviour of the shorter distance together make it suitable for the study of the spectroscopy of bottomonium. Additionally, we have also incorporated the spin dependent part of one gluon exchange potential. The detailed description of the form of instanton potential for heavy quark can be found in \cite{Yakhshiev2017} where authors have added the spin dependent attributes into it. This potential has a huge history \cite{Wilczek1978, Callan1978, Eichten1981} and its central part was  derived long ago \cite{Diakonov1989} based on the instanton liquid model \cite{Diakonov1984,Diakonov1986,Diakonov2003} for the QCD Vacuum. With the use of instanton potential acting between quark and anti-quark pair, we solved the non-relativistic schrodinger equation by variational method. To test the predictivity of the potential we have further calculated various decay properties like vector and pseudoscalar decay constants, di-leptonic, di-gamma, di-gluon as well as tri-gluon decays together with radiative transition decay widths. The computed results are tabulated and compared with other available theoretical and experimental data and finally we draw our own conclusions.

\section{Theoretical Framework to compute mass spectra}
Evaluation of the bound state mass spectra demands the solution of the Schr\"{o}dinger equation with the potential from the instanton vacuum. Unfortunately, the Schr\"{o}dinger equation is not exactly solvable for most of the systems. So, generally one adopt approximate  methods like perturbation, variational or WKB methods. In the present work, the variational method is employed to calculate the ground and excited state mass spectra. For the treatment of bottomonium on the non-relativistic footing the Hamiltonian can be written as
\begin{equation}
H = \frac{P^2}{2M} + V\left(r\right)
\end{equation}
Here, $P$ is the relative momenta, $M$  represents reduced mass , $V(r)$ is the potential acting between quark and anti-quark. 

Now as described earlier, the form of the potential we have used is instanton induced potential according to the Instanton Liquid Model (ILM)\cite{Yakhshiev2017}. The ILM has two important parameters portraying the diluteness of the instanton liquid \cite{Shuryak1982, Diakonov1984} , one is the average size of instanton $\bar{\rho}$ and other is the average distance between instantons $\bar{R}$. Numerical values of these parameters are different for different approches. For example, Shuryak \cite{Shuryak1982} proposed the values of these parameters as $\bar{\rho}\simeq 0.33 fm$ and $\bar{R}\simeq 1fm$. In ref \cite{Kim2006, Goeke2007, Goeke20072} , $\bar{\rho}\simeq 0.35 fm$ and $\bar{R}\simeq 0.856fm$ were used in their $1/N_c$ meson loop contibution for the light quark sector. And in this framework the range of the potential is identified only by the size of the instanton $(\bar{\rho})$.

For $r\ll \bar{\rho}$ i.e., when the distance between quark-antiquark is smaller than the average size of instanton, the central potential is given by \cite{Yakhshiev2017} %the dimensionless integral can be expanded interms of $x$ as \cite{Yakhshiev2017}
%\begin{eqnarray}\label{Ix}
%{I(x)}&\simeq &\left[{\frac{\pi^3}{48}}-{\frac{\pi^3}{3}{J_1\left(2 %\pi\right)}}\right]{x^2} \nonumber \\ && +  %\left[-\frac{{\pi^3}{\left(438+7\pi^2\right)}} {30720} + {\frac{{J_2}{\left({2 %\pi}\right)}}{80}}\right]{x^4} + \mathcal{O}\left(x^6\right)
%\end{eqnarray}
\begin{equation}\label{eq:b}
{V(r)}\simeq\frac{4  {\pi} \bar{\rho}^3 }{{\bar{R^4}}{N_c}}\left(1.345{\frac{r^2}{\bar{\rho}^2}}-0.501{\frac{r^4}{\bar{\rho}^4}}\right)
\end{equation}
here, $N_{c}=3$ represents the colour degrees of freedom.\\
For $r\gg\bar{\rho}$ i.e., the distance between quark-antiquark is higher than the size of instanton, the central part of the potential is given by \cite{Yakhshiev2017} %the dimensionless integral becomes
%\begin{equation}\label{eq:c}
%{I(x)}\simeq {-\frac{2\pi^2}{3}} \left[\pi J_0 \left(\pi\right)+ J_1 \left(\pi\right)\right]{-\frac{\pi^2}{2 x}} + \mathcal{O}\left(x^{-2}\right)
%\end{equation}
\begin{equation}\label{eq:c2}
{V(r)}\simeq 2 \Delta M_Q - {\frac{g_{np}}{r}}
\end{equation}
here, $\Delta M_Q$ is the correction to heavy-quark mass from the instanton vacuum and is given by \cite{Yakhshiev2017}.
\begin{equation}\label{eq:c3}
{\Delta M_Q} ={-\frac{4 \pi^4 \bar{\rho}^3 }{3 \bar{R^4} N_c}} \left( J_0 \left(\pi\right)+ \frac {1}{\pi} J_1 \left(\pi\right)\right)
\end{equation}
Here, $J_0$ and $J_1$ are the Bessel functions. The coupling constant in equation $(4)$ is defined as $g_{np}= 2 \pi^{3} {\bar{\rho}^4} / N_c {\bar{R}}^4 $ and can be considered as nonperturbative correction to the strong coupling constant ${\alpha}_s (r)$. When $r$ reaches to infinity, the potential is saturated at $2\Delta M_{Q}$ and this signifies that instanton vacuum cannot explain quark confinement \cite{Diakonov1989}. Thus for the present study, we have added a  state dependent confinement potential $V_0$ which has the form
%\begin{equation}\label{eq:c}
%{V_0\left(n+l\right)} = -a \left(n+l\right)^4 + b \left(n+l\right)^3 - c \left(n+l\right)^2 + d \left(n+l\right)+ e
%\end{equation}
\begin{equation}\label{eq:c4}
{V_0\left(z\right)} = \left(\ln \left(z\right)+a\right)b 
\end{equation}
Where, $z = 4n+3l-1 $ and the numerical values of the constants are $a$ = 0.72 MeV and $b$ = 290.4 MeV. These are our fitted model parameters which are deduced by fitting the masses of low lying bottomonium states based on instanton potential framework. For instanton liquid model framework, potential is defined only for $r\ll \bar{\rho}$ and $r\gg \bar{\rho}$ and to remove the discontinuity at $r = \bar{\rho}$ we have further added $V_{1}$ = 61.81 MeV into eq. $(3)$. This inclusion changes the potential range from $r\ll \bar{\rho}$ to $r\leq \bar{\rho}$ and $r\gg \bar{\rho}$ to $r\geq \bar{\rho}$.

%\begin{equation}\label{eq:d}
%{V_{SS}\left(r\right)}= \frac{1}{{3}m_{Q}^{2}}  \nabla^2 {V_c\left(r\right)}
%\end{equation}

%\begin{equation}\label{eq:e}
%{V_{LS}\left(r\right)}= \frac{1}{{2}m_{Q}^{2}}  \frac{1}{r} %\frac{{d}{V_c\left(r\right)}}{dr}
%\end{equation}

%\begin{equation}\label{eq:f}
%{V_{T}\left(r\right)}= \frac{1}{{3}m_{Q}^{2}} \left( \frac{1}{r} %\frac{{d}{V_c\left(r\right)}}{dr}-\frac{{d^2}{V_c\left(r\right)}}{dr^2} \right)
%\end{equation}

%\begin{equation}\label{eq:d}
%\phi_{n}^{L}\left({r}\right) =  {\left(\frac{2^{2L+\frac{7}{2}} %r_{n}^{-2L-3}}{\sqrt{\pi}\left({2L+1}\right)!!\right)^{\frac{1}{2}}
%\end{equation}

%\begin{equation}\label{eq:d}
%\phi_{n}^{L}\left({r}\right)=\left(2^{2L+{\frac{7}{2}}} {r_{n}^{-2L-3}}\right)
%\end{equation}

%\begin{equation}\label{eq:d}
%\phi_{n}^{L}\left({r}\right)={\left(\frac{2^{2L+{\frac{7}{2}}} {r_{n}^{-2L-3}}}{\sqrt{\pi} \left({2L+1}\right)!!}\right)}^{\frac{1}{2}}{r^L} {\exp^{-\left(\frac{r}{n}\right)}}^2
%\end{equation}
%\section{Leptonic and Digamma Decays of $B$ and $B_s$ Meson}\label{hadronic}

The spin average mass of S wave $b\bar{b}$ system can be obtain as 
% To compute experimental spin average masses pseudoscalar and the vector masses listed by \cite{PDG2018} are used in the following relation,
\begin{equation}\label{eq:c1}
M_{SA} = m_b + m_{\bar{b}} + \langle {H} \rangle
\end{equation}
For the S wave further hyperfine split will give vector and pseudoscalar components.

\begin{equation}\label{vss}
V^{SS}_{Q \bar Q} (r) = \frac{8}{9} \frac{\alpha_s} { m_{Q} m_{\bar{Q}}} \vec{S_Q}. \vec S_{\bar Q}  {4\pi} {\delta^3 (r)}
\end{equation}

%To the P and D wave spin average mass we have incorporated the %contribution of the spin-orbit and tensor part through one %gluon exchange potential (OGEP) \cite{PCV1992}.

For the massess of $P$ and $D$ waves, we have incorporated the contribution of the spin-orbit and tensor part of the one gluon exchange potential (OGEP) of the form given by \cite{PCV1992, Mshah2016e, Mshah2014, Mshah2016, Monterio2010, Bhavsar2018}

\begin{eqnarray}\label{Vls}
V^{LS}_{Q \bar Q} (r) &=& \frac{\alpha_s}{4} \frac{N_Q^2 N_{\bar Q}^2}{(E_Q + m_{Q})(E_{\bar{Q}} + m_{\bar{Q}})}  \frac{\lambda_Q . \lambda_{\bar Q}}{2 \ r} \\ &&\otimes \left[ \left[ \vec{r} \times (\hat{p_Q}-{\hat p}_{\bar{Q}}).(\sigma_Q + \sigma_{\bar Q})\right]\left( {D'_0 (r)}+ 2 {D'_1 (r)} \right) \right. \nonumber\\ &&  \left. + \left[ \vec{r} \times (\hat{p_Q}+ {\hat p}_{\bar Q}). (\sigma_i - \sigma_j)\right]\left( {D'_0 (r)}-  {D'_1 (r)} \right) \right]  \nonumber
\end{eqnarray}
and
\begin{eqnarray}\label{Vt}
V^{T}_{Q \bar Q} (r) &=& - \frac{\alpha_s}{4} \frac{N_Q^2 N_{\bar Q}^2}{(E_Q + m_{Q})(E_{\bar{Q}} + m_{\bar{Q}})} \nonumber \\ && \otimes \ \lambda_Q . \lambda_{\bar Q} \left( \left( \frac{D''_1 (r)}{3}- \frac{D'_1 (r)}{3 \ r} \right) S_{Q \bar Q}\right)
\end{eqnarray} \\

where $S_{Q \bar Q} = \left[ 3 (\sigma_Q. {\hat{r}})(\sigma_{\bar Q}. {\hat{r}})- \sigma_Q . \sigma_{\bar Q}\right]$ with ${\hat{r}} = {\hat{r}}_Q - {\hat{r}}_{\bar Q}$ as the unit vector in the direction of $\vec{r}$.\\

Expression for the confined gluon propagators are taken from \cite{PCV1992} as,

\begin{equation}
D_0 (r) = \left( \frac{\alpha_1}{r}+\alpha_2 \right) \exp(-r^2 c_0^2/2)
\end{equation}

\begin{equation}
D_1 (r) =  \frac{\gamma}{r} \exp(-r^2 c_1^2/2)
\end{equation}

For the present study we have adopted the same values of the parameters, $\alpha_1=1.035$, $\alpha_2=0.3977$, $c_0 = 0.3418$ GeV, $c_1=0.4123$ GeV, $\gamma=0.8639$ as used by \cite{Monterio2010}. And the instanton potential parameters as $\bar{\rho}\simeq 0.36 fm$ and $\bar{R}\simeq 0.96 fm$ are being used. \\

The strong running coupling constant $\alpha_{s}$ calculated as
\begin{equation}
\alpha_{s}  \left( M^{2} \right) = \frac{4\pi}{\left(11-\frac{2}{3}n_f\right) \left(\ln \frac{M^2}{\Lambda^2} \right)}
\end{equation}
where, $\Lambda_{QCD}$ is taken as 0.156 GeV which will give the PDG \cite{PDG2018} listed value of $\alpha _s$ at the $Z$ meson mass (91 GeV) range of 0.0118. Employing $\Lambda_{QCD}$ =0.156 GeV and putting spin average mass (M) of respective state in above equation gives $\alpha_s$ for corresponding state. Here, $n_f$ represents the number of flavours and for bottom sector it is 4.

The harmonic oscillator trial wave-function used here is
\begin{equation}
R_{nl}\left({r}\right)= {N_{nl}}                     {\left(\mu{r}\right)}^{l}  \exp\left({-\mu^{2}{r^2}/2}\right)    \L_{n-1}^{l+1/2} {\left(\mu^2 r^2\right)}
\end{equation}
with $N_{nl}$ being the normalization constant expressed as
\begin{equation}
N_{nl}=\sqrt{\frac{2\mu^3 \left(n-1\right)!}{\Gamma\left(n+l+1/2\right)}}
\end{equation}
Also, $\mu $ is the variational parameter and ${L_{n-1}^{l+1/2}}{(\mu^2 r^2)}$ is the associated Lagurre polynomial.

Based on the above theoretical formalism, we have computed the spectroscopic masses of the $S$, $P$ and $D$ wave bottomonium using instanton induced potential between the quark and anti-quark. The spin averaged masses of the radial $(nS)$ and orbital excited states $(nP,nD)$ are computed and further spin dependent one gluon exchange interactions are used to remove the mass degeneracy of the bottomonium states. Results obtained here are tabulated in the Table \ref{tab1} for $S$ wave masses and in Table \ref{tab2} for $P$ and $D$ wave masses. We have compared our data set with the theoretical model predictions such as relativistic Dirac Model \cite{Bhavsar2018}, QCD Relativistic functional approach \cite{Fischer2015} where authors have used two different methods rainbow-ladder truncation of Dyson-Schwinger and Bethe-Salpeter  in search of the effects of the varying shapes of the effective running coupling on ground as well as excited states in the channels having quantum numbers J less than or equal to 3, Constituent quark model\cite{Segovia2016}  , Relativistic quark model \cite{Godfrey2015, Ebert2011}, Relativistic potential model\cite{Radford2011} and with the recent experimental data listed by PDG\cite{PDG2018}. Also, Figure 1 shows the energy level diagram of the bottomonium states where we have compared our predictions with the PDG \cite{PDG2018}reported values.

\begin{table*}[htp]
%\begin{center}
\tabcolsep 7pt
 \small
%\caption{S-wave $B$ ($b\bar{q}$, $q \in u, d$) spectrum (in MeV).} \label{tab1}
\caption{S-wave mass spectrum of $ b \bar{b}$  bound states (in MeV).} \label{tab1}
\begin{tabular}{cccccccccccccccc}
\hline\hline
% &    &  &       &&    & \multicolumn{2}{c} {PDG} &&& & && & \\
%\cline{7-8}
%nS & $J^{PC}$ & State & Present & $\langle V_{Q \bar q}^{j_1j_2}\rangle$ &Present  &  Meson  & Mass\cite{PDG2012} & \cite{zeng1995} &\cite{Ebert2010} & \cite{Devlani2012} & \cite{Pierro} & \cite{Lahde2000} & \cite{Dowdall2012}
%& \cite{Joachim2000} \\

\hline
%&    &         \multicolumn{3}{c} {Present} &&& & && & & &\\
%\cline{3-4}
nS &  State & $J^{PC}$ & Present & PDG \cite{PDG2018}\footnote{Particle Data Group 2018} & \cite{Bhavsar2018}\footnote{Relativistic Dirac model} & \cite{Fischer2015}\footnote{Relativistic functional approach} & \cite{Segovia2016}\footnote{Constituent quark model} & \cite{Godfrey2015} \footnote{Relativistic quark model}& \cite{MShah2012}\footnote{Non-relativistic potential model}& \cite{Ebert2011}\footnote{Relativistic quark model} & \cite{Radford2011}\footnote{Relativistic potential model}\\

\hline

1S & $1{^3S_1}$ & $1^{--}$  & 9460.75 &  9460.30 $\pm$ 0.26  & 9460.99  & 9490  & 9502 & 9465 & 9460 & 9460 & 9460 \\
   & $1{^1S_0}$  & $0^{-+}$  & 9412.22 &   9399.0 $\pm$ 2.3& 9390.7 & 9414 & 9455  & 9402 & 9392  & 9398  & 9393 \\

2S & $2{^3S_1}$ & $1^{--}$  & 10026.22  &    10023.26 $\pm$ 0.31  & 10024.1 & 10089 & 10015 & 10003 & 10024 & 10023 & 10023\\
   & $2{^1S_0}$ & $0^{-+}$  & 9995.48   &            \ldots   & 9999.3& 9987 & 9990 & 9976 & 9991  & 9990 & 9987\\

3S & $3{^3S_1}$ & $1^{--}$  & 10364.65    &        10355.2 $\pm$ 0.5        & 10356.2 & 10327 & 10349 & 10354 & 10346 & 10355 & 10364 \\
   & $3{^1S_0}$ & $0^{-+}$  & 10339.00    &     $\cdots$         &  10325.3 & \ldots & 10330 & 10336 & 10323 &  10329 & 10345\\

4S & $4{^3S_1}$ & $1^{--}$  &  10594.47  &         10579.4 $\pm$ 1.2 & 10576.2 & \ldots  & 10607 & 10635  & 10575 &  10586 & 10643 \\
   & $4{^1S_0}$ & $0^{-+}$  & 10572.49  &         $\cdots$          & 10554.4 & \ldots & \ldots & 10523& 10558 & 10573 & 10364 \\

5S & $5{^3S_1}$ & $1^{--}$  & 10766.14   &  10889.9        & 10758.5  &\ldots  & 10818 & 10878 & 10755 & 10869 & \ldots\\
   & $5{^1S_0}$ & $0^{-+}$  &  10746.76   &        $\cdots$          & 10738.4 &  \ldots& \ldots & 10869 & 10741 & 10851& \ldots\\

6S & $6{^3S_1}$ &  $1^{--}$  & 11081.70    &  10992.9        & \ldots &\ldots  & 10995 & 11243 & 10904 & 11088 & \ldots\\
    & $6{^1S_0}$ &  $0^{-+}$  & 11064.47    &        $\cdots$          & \ldots  &  \ldots& \ldots & 11226 & 10892 & 11061 & \ldots\\
\hline
\hline

\end{tabular}

%\end{center}

\end{table*}

\section{Decay properties of heavy quarkonia $Q\bar{Q}$}

For any potential model, apart from the mass spectra other observables like radiative decays and annihilation decays are important testing ground to know the inter-quark interactions. Keeping this view, we have computed the decay properties of $b\bar{b}$ states with no additional parameters.
Further, We have incorporated these decay properties with QCD correction whenever it is possible. 

\subsection{Pseudoscalar and vector decay constants}

Estimation of the decay constants of mesons constituting heavy quarks is essential part of the study as it offers the information of the CKM (Cabibbo-Kobayashi-Maskawa) matrix elements.
The conventional formula within the non-relativistic limit for the pseudoscalar and vector states of the $S$ wave heavy quarkonia is the Van Royan-Weiskopff formula \cite{VanRoyen}. The Van Royan- Weiskopff formula was derived  under the limit that the spinors of the quark antiquark inside the mesonic system are approximated to two component Pauli spinors \cite{VanRoyen}. This formula relates ground state wave function at the origin to the decay constant and is expressed as
%\begin{equation}\label{eq:d}
%f_{P/V}{^2}= 3 %\frac {{{{R_{{ns}{P/V}} \left(0\right) }}}{\pi{M_{{ns}{P/V}} }
%{f_{P/V}{^2}}= 3 \frac{|R_{{ns}{P/V}}\left(0\right)|}}{\pi{M_{{ns}{P/V}}}} {{\bar{C}}^2 \left(\alpha s\right)} |R|
%\end{equation}
\begin{equation}
{\textit{f}}^2_{P/V} \left(nS\right)= \frac{3|{R_{ns}\left(0\right)|}^2}{\pi M_{ns}} {{\bar{c}^2}\left(\alpha s\right)}
\end{equation}
Where, the first order QCD corrections factor $\bar{c}^2\left(\alpha s\right)$ is given by \cite{Braaten1995, Berezhnoy1996}
\begin{equation}\label{eq:d}
{\bar{c}}\left(\alpha s\right)= 1-\frac{\alpha_{s}}{\pi} \left(\delta^{P,V}-{\frac {m_1-m_2}{m_1 + m_2}\ln \frac{m_1}{m_{2}}}\right)
\end{equation}
%{{\bar{C}}^2 \left(\alpha s\right)}= 1 {\plus} {\frac{\alpha {s}}{\pi}}
%\left( \frac{m_Q \minus m_{\bar{Q}}}{m_Q \plus m_{\bar{Q}}}\right)

In the case of quarkonia, the second term inside the bracket in equation (16) vanishes and $\delta^{V}$=8/3 in the case of vector state and $\delta^{P}$=2 for the pseudoscalar state. The calculated the pseudoscalar and vector decay constants with and without the first order QCD corrections are presented in Table \ref{tab3} and \ref{tab4} and are compared with other available model predictions.

\begin{figure}[tbp]
\centering
%\resizebox{0.35\textwidth}{!}{%
\includegraphics[width=.50\textwidth]{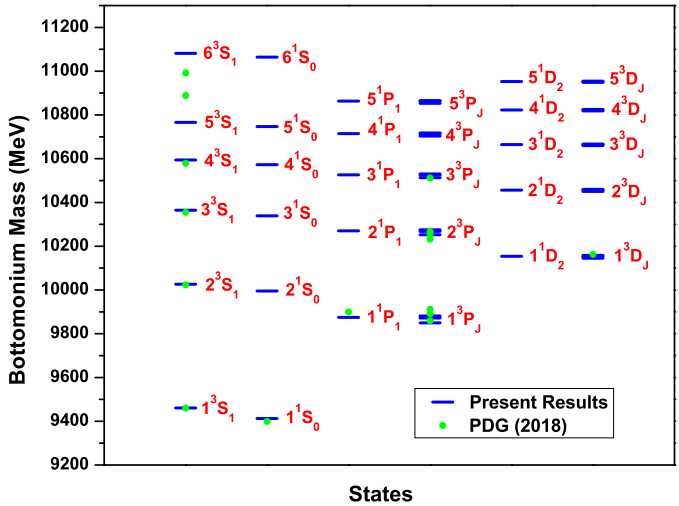}
\caption{$ b  \bar{b}$ mass spectrum }\label{mixing2}
\end{figure}

\begin{table*}[htp]
\begin{center}
\tabcolsep 7pt
 \small
\caption{P-wave and D-wave mass spectra of $b\bar{b}$  (in MeV)} \label{tab2}
\begin{tabular}{ccccccccccccccc}
\hline\hline
% &  &   &   &&&&           & \multicolumn{2}{c} {Experiment} & &&&&\\%
%\cline{9-10}%
nL &  State & $J^{PC}$ & Present    & PDG \cite{PDG2018}  & \cite{Bhavsar2018} &\cite{Fischer2015} & \cite{Segovia2016} & \cite{Godfrey2015} & \cite{MShah2012}  & \cite{Ebert2011}   & \cite{Radford2011} & \cite{Li2009} \\
\hline\
1P & $1{^3P_2}$ & $2^{++}$  & 9881.40 &  9912.21 $\pm$ 0.26   & 9912.3 & 9906 & 9886 & 9897 & 9908 & 9912 & 9912 & 9918 \\
   & $1{^3P_1}$ & $1^{++}$  & 9871.47 &  9892.78 $\pm$ 0.26  & 9901.8  & 9842 & 9874& 9876 & 9888 & 9892 & 9891 & 9897\\
   & $1{^3P_0}$ & $0^{++}$  & 9849.61 &  9859.44 $\pm$0.42   & 9889.2 & 9815 & 9855 & 9847 & 9862 & 9859 & 9861 & 9865\\

   & $1{^1P_1}$ & $1^{+-}$  & 9874.56 &  9899.3$\pm$ 0.8 & 9854.1 & 9806 & 9879& 9882 & 9896 & 9900 & 9900 & 9903\\

2P & $2{^3P_2}$ & $2^{++}$   & 10274.77 &   10268.65$\pm$ 0.22   & 10265.9 & \ldots & 10246& 10224 & 10268 & 10268 & 10271 & 10269 \\
   & $2{^3P_1}$ & $1^{++}$   &  10267.86 &  10255.46 $\pm$ 0.22    & 10258.9 & 10120  & 10236 & 10246 & 10256 & 10255 & 10255 & 10251\\
   & $2{^3P_0}$ & $0^{++}$   & 10252.54 &     10232.50$\pm$ 0.40     & 10234.7 & 10254 & 10221 & 10226& 10241 & 10233 & 10230 & 10226 \\
   & $2{^1P_1}$  & $1^{+-}$   & 10270.00 &  \ldots   & 10264.9 & 10154 & 10240 & 10250 & 10261 & 10260  & 10262 & 10256\\

3P & $3{^3P_2}$  & $2^{++}$  & 10530.21 & \ldots& 10516.9 & \ldots & 10521 & 10550 & 10516 & 10550 & \ldots & 10540 \\
   & $3{^3P_1}$ & $1^{++}$  & 10524.84 & 10512.1$\pm$ 2.3& 10508.8 & 10303& 10513 & 10538  & 10507 & 10541 &\ldots & 10529 \\
   & $3{^3P_0}$ & $0^{++}$  & 10512.88 & \ldots  & 10497.6  & \ldots& 10500 & 10552 & 10511 & 10521 & \ldots & 10502\\
   & $3{^1P_1}$ & $1^{+-}$  & 10526.50 & \ldots & 10540.2 & \ldots & 10516& 10541 & 10497 & 10544 & \ldots & 10529  \\

4P & $4{^3P_2}$  & $2^{++}$  & 10717.86 & \ldots& 10707.0 & \ldots & 10798 & 10550 & \ldots & 10812& \ldots & 10767 \\
   & $4{^3P_1}$ & $1^{++}$  & 10713.44 & \ldots& 10706.5 & \ldots& 10788 & 10538  & \ldots & 10802 &\ldots & 10753 \\
   & $4{^3P_0}$ & $0^{++}$  & 10703.56 & \ldots  & 10703.8  & \ldots& 10781 & 10552 & \ldots & 10781 & \ldots & 10732\\
   & $4{^1P_1}$   & $1^{+-}$  & 10714.80 & \ldots & 10704.6 & \ldots & 10790& 10541 & \ldots & 10804 & \ldots & 10757  \\

1D & $1{^3D_3}$ & $3^{--}$  &   10158.31 &   $\cdots$   & 10140.4  & 10232 & 10127 & 10115 & 10177 & 10166 & 10163 & 10156\\
   & $1{^3D_2}$ & $2^{--}$  &  10152.77 &  10163.7$\pm$ 1.4  & 10138.7& 10145& 10122 & 10147 & 10162 & 10161 & 10157 & 10151\\
   & $1{^3D_1}$ & $1^{--}$  &  10144.99 &  $\cdots$   & 10136.0& \ldots & 10117& 10138 & 10147 & 10154 & 10149 & 10145\\
   & $1{^1D_2}$  & $2^{-+}$  & 10153.80 &  $\cdots$ & 10068.2  & 10194  & 10123  & 10148 & 10166 & 10163 & 10158 & 10152\\

2D & $2{^3D_3}$ & $3^{--}$  & 10459.85 &    $\cdots$  & 10398.7 & \ldots & 10422 & 10455 & 10447 & 10449 & 10456 & 10442\\
   & $2{^3D_2}$ & $2^{--}$  & 10455.86 &  $\cdots$  & 10397.1& \ldots & 10418 & 10449 & 10437 & 10443 & 10450 & 10438\\
   & $2{^3D_1}$ &  $1^{--}$  & 10450.23 &    $\cdots$  & 10395.7 & \ldots & 10414 & 10441 & 10428 & 10435 & 10443 & 10432\\
   & $2{^1D_2}$ &  $2^{-+}$  &  10456.60 &    $\cdots$  & 10336.0 & \ldots & 10419& 10450 & 10440 & 10445 & 10452 & 10439\\

3D & $3{^3D_3}$ &  $3^{--}$  & 10667.25 & \ldots & 10620.9  & \ldots & \ldots & 10711  & 10652 & 10717 & \ldots & 10680\\
   & $3{^3D_2}$ & $2^{--}$  & 10664.12 & \ldots & 10619.3 & \ldots& \ldots& 10705& 10645& 10711 & \ldots & 10676\\
   & $3{^3D_1}$ & $1^{--}$  & 10659.68 & \ldots & 10616.8  & \ldots & \ldots & 10698 & 10637 & 10704 & \ldots & 10670\\
   & $3{^1D_2}$ & $2^{-+}$  & 10664.70 & \ldots& 10564.3 & \ldots  & \ldots& 10706 & 10646 & 10713n & \ldots & 10677\\

4D & $4{^3D_3}$ &  $3^{--}$  & 10825.12  & \ldots & 10820.9 & \ldots & \ldots & 10939 & 10817 & 10963 & \ldots & 10886\\
   & $4{^3D_2}$ & $2^{--}$  & 10822.52 &  \ldots & 10819.3 & \ldots& \ldots & 10934 & 10811 & 10957 & \ldots & 10882\\
   & $4{^3D_1}$ & $1^{--}$  & 10818.83  &  \ldots  & 10816.9 & \ldots & \ldots & 10928 & 10805 & 10949 &\ldots & 10877\\
   & $4{^1D_2}$ & $2^{-+}$  & 10823.00  & \ldots  & 10768.8 & \ldots & \ldots  & 10935 & 10813 & 10959 & \ldots & 10883 \\

5D & $5{^3D_3}$ & $3^{--}$  & 10954.42 & \ldots & 11005.2 & \ldots & \ldots & \ldots & 10955 & \ldots & \ldots & 11069\\
   & $5{^3D_2}$ & $2^{--}$  & 10951.59 &  \ldots & 11003.7 & \ldots& \ldots & \ldots & 10950 & \ldots & \ldots & 11065\\
   & $5{^3D_1}$ &  $1^{--}$  & 10949.01  &  \ldots  & 11001.4 & \ldots & \ldots & \ldots & 10945 & \ldots &\ldots & 11060\\
   & $5{^1D_2}$ & $2^{-+}$  & 10952.60  & \ldots  & 10956.7 & \ldots & \ldots  & \ldots & 10952 & \ldots & \ldots & 11066 \\
    \hline\hline
 \end{tabular}

\end{center}
\end{table*}

\subsection{The Leptonic Decays of bottomonium states}

%The $1^{\minus\minus}$ states decays leptonically via pair of virtual photon.
Descriptive study of leptonic decays of mesons not only provide underlying dynamics of quark anti-quark annihilation \cite{Bhatnagar2006} but also used as a probe to study compactness of the pair of quark-antiquark \cite{Rosner2006}. In addition to that leptonic decay rates can also help us to distinguish conventional mesons and exotic states \cite{Badalian1987} .

The vector mesons $(1^{--})$ annihilates leptonically via single virtual photon and leptonic partial decay width computed using Van Royen- Weisskopf formula \cite{VanRoyen} along with the QCD correction of one loop level given as \cite{Kwong1988, Barbieri1975}
%\begin{equation}\label{eq:d}
%\Gamma\left({n^3} {s_1}\right)= 4 {\alpha^2} {e_b^2} |R_n\left(0\Right)
%\end{equation}
\begin{eqnarray}\label{eq:af}
\Gamma(n^3 S_1\rightarrow l^+ l^-)=\frac{4 \alpha_e^2 {e_Q^2}}{M_V^2} |R_{nl} \left(0\right)|^2 \left[1-\frac{16 \alpha_s}{3 \pi}\right] \ \ \ \ \
\end{eqnarray}

Here, $ |R_{nl}(0)|^{2}$ is the square of the radial wave function at the origin. $\alpha_{e} =1/137$ and $\alpha_{s}$ are the electromagnetic coupling constant and strong running coupling constant respectively. $e_Q$ is the charge of the heavy quark in units of electron charge.
For the $D$ wave states $n^3 D_1$ leptonic decay width can be given as \cite{Novikov1978},

\begin{eqnarray}\label{eq:af}
\Gamma(n^3 D_1\rightarrow l^+ l^-)=\frac{25 \alpha_e^2 {e_Q^2}}{2 m_{b}^4  {M^2\left(n^3 D_1\right)}} |R_{nl}^{''}\left(0\right)|^2 \left[1-\frac{16 \alpha_s}{3 \pi}\right]  \ \ \ \ \
\end{eqnarray}

For the $n^3D_1$ bound states annihilation  into $l^+l^-$ have same first order QCD correction factor as of $n^3S_1$ bound states \cite{Bradley1981}. The leptonic decay widths are tabulated in Table \ref{tab5} with and without QCD corrections. We have compared our results with other available experimental as well as theoretical data.

\begin{table*}[htp]

\tabcolsep 10pt
 \small
\caption{The vector decay constant(in MeV) of the $S$ wave bottomonium
 states.} \label{tab3}
\begin{tabular}{ccccccccccccccc}
\hline\hline
% &  &   &   &&&&           & \multicolumn{2}{c} {Experiment} & &&&&\\%
%\cline{9-10}%

\hline
 State & $J^{PC}$  & $\textit{f}_V$ & $\textit{f}_V (R) $ & PDG \cite{PDG2018}   & \cite{Colquhoun2015} &\cite{Bhavsar2018} & \cite{BPatel2009} & \cite{Wang2006} & \cite{Bhaghyesh2011}\\
\hline
$1{^3S_1}$ & $1^{--}$  & 653.44  & 551.53 & 715 $\pm$ 5 & 649(31)  & 705.4 & 706  & 498 $\pm$ (20)& 831  \\

$2{^3S_1}$ & $1^{--}$  & 563.73 & 477.05 & 498 $\pm$ 8 & 481(39)& 554.9  &   547 & 366 $\pm$ (27)& 566 \\

$3{^3S_1}$ & $1^{--}$  & 507.90 & 430.42 & 430 $\pm$ 4 & \ldots & 436.8 &  484 & 304 $\pm$ (27) & 507  \\

$4{^3S_1}$& $1^{--}$  &  466.60 & 395.80  & 336 $\pm$18 & \ldots & 332.4 & 446 & 259 $\pm$ (22) & 481  \\

$5{^3S_1}$ & $1^{--}$  &  434.61   & 368.91    & \ldots       & \ldots    & 286.5  & 419 & 228 $\pm$ (16)  & 458  \\

$6{^3S_1}$ & $1^{--}$  &  406.39   &   345.40  & \ldots       & \ldots    &   &  &  &   \\

\hline\hline
\end{tabular}
\end{table*}

\begin{table*}

\tabcolsep 7pt
 \small
\caption{The pseudoscalar decay constant(in MeV) of the $S$ wave bottomonium
 states.} \label{tab4}
\begin{tabular}{ccccccccccccccc}
\hline\hline
% &  &   &   &&&&           & \multicolumn{2}{c} {Experiment} & &&&&\\%
%\cline{9-10}%

\hline
State  & $J^{PC}$ &  $\textit{f}_P$ & $\textit{f}_P (R) $ & \cite{Soni2018} & \cite{Krassnigg2016} & Lattice \cite{Davies}  & \cite{Bhaghyesh2011} & QCD sum rules \cite{Veli2012} & \cite{Negash2016}  \\
\hline
$1{^1S_0}$ & $0^{-+}$  & 654.81  & 578.21 & 646.025 & 756  & 667 &   834& 251$\pm$ 0.072 & 1016.8  \\
$2{^1S_0}$ & $0^{-+}$   & 564.59  & 499.48 & 518.803 & 285  & \ldots &  567&  \ldots & 806.6  \\
$3{^1S_0}$ & $0^{-+}$  & 508.53  & 450.35 & 474.954 & 333  & \ldots  & 508 & \ldots& 713.4  \\
$4{^1S_0}$ & $0^{-+}$  & 467.09  & 413.93 & 449.654 & 40  &  \ldots &   \ldots& \ldots & \ldots   \\
$5{^1S_0}$ & $0^{-+}$  & 435.00  & 385.68 & 432.072 & \ldots  & \ldots  & \ldots & \ldots & \ldots   \\
$6{^1S_0}$ & $0^{-+}$  & 406.71  & 360.93 &  & \ldots  & \ldots  & \ldots & \ldots & \ldots   \\

\hline\hline
\end{tabular}
\end{table*}

\begin{table*}
\tabcolsep 7pt
 \small
\caption{The di-leptonic decay widths (in keV) of the $S$ wave and $D$ wave bottomonium
 states.} \label{tab5}
\begin{tabular}{ccccccccccccccc}
\hline\hline
% &  &   &   &&&&           & \multicolumn{2}{c} {Experiment} & &&&&\\%
%\cline{9-10}%

\hline
 State  & $J^{PC}$ & $\Gamma_{{\textit{l}^+}{\textit{l}^-}}$ & $\Gamma_{{\textit{l}^+}{\textit{l}^-}}$ (R) &    PDG \cite{PDG2018}  & \cite{Radford2011} &\cite{Segovia2016} & \cite{MShah2012} & \cite{Bhavsar2018} & \cite{PCV1992} & \cite{Ebert2003} & \cite{Li2009} \\
\hline
$1{^3S_1}$ & $1^{--}$  & 1.1191 &  0.7700 & 1.34 $\pm$ 0.018 & 1.33  & 0.71 & 1.20  & 1.30 & 1.809 & 1.3 & 1.60 \\

$2{^3S_1}$ & $1^{--}$  & 0.7859 & 0.5442   & 0.612 $\pm$ 0.011 & 0.62 & 0.37 & 0.52  & 0.76 & 0.797 & 0.5 & 0.64\\

$3{^3S_1}$ & $1^{--}$  &0.6171 & 0.4288 &  0.443 $\pm$ 0.008 & 0.48 & 0.27 & 0.33 & 0.45 & 0.618 &\ldots & 0.44\\

$4{^3S_1}$ & $1^{--}$  & 0.5096 & 0.3549 &  0.272 $\pm$ 0.029 & 0.40 & 0.21 & 0.24& 0.26 & 0.541 & \ldots  & 0.35\\

$5{^3S_1}$ & $1^{--}$  & 0.4350  & 0.3035      & 0.31 $\pm$ 0.07    & \ldots    & 0.18 & 0.19& 0.18 &  \ldots &  \ldots & 0.29 \\

$6{^3S_1}$ & $1^{--}$  & 0.3695  & 0.2586      &     & \ldots    &  & &  &  \ldots &  \ldots &  \\

$1{^3D_1}$ & $1^{--}$  &  0.0072     & 0.0050       &  \ldots       &  \ldots     &  \ldots &  \ldots & 0.106 &  \ldots  &  \ldots &  \ldots\\

$2{^3D_1}$ & $1^{--}$  & 0.0084      &   0.0058     &   \ldots      &   \ldots    &  \ldots &  \ldots & 0.078 & \ldots  &  \ldots &  \ldots \\
$3{^3D_1}$ & $1^{--}$  &  0.0085    &   0.0059     &  \ldots       &   \ldots    &  \ldots &   \ldots& 0.051 & \ldots  &  \ldots &  \ldots \\
$4{^3D_1}$ & $1^{--}$  &  0.0084     &  0.0058    &   \ldots      &  \ldots     &   \ldots &  \ldots & 0.042 & \ldots  &  \ldots &  \ldots\\
$5{^3D_1}$ &  $1^{--}$  & 0.0082    &   0.0057     &   \ldots      &  \ldots     &  \ldots  &  \ldots & 0.028 & \ldots  &  \ldots &  \ldots\\
\hline\hline
\end{tabular}
\end{table*}

\begin{table*}
\tabcolsep 10pt
 \small
\caption{The di-gamma decay widhts (in keV) of the $S$ wave and $P$ wave bottomonium
 states.} \label{tab6}
\begin{tabular}{ccccccccccccccc}
\hline\hline
% &  &   &   &&&&           & \multicolumn{2}{c} {Experiment} & &&&&\\%
%\cline{9-10}%

\hline 
 State  & $J^{PC}$ & $\Gamma_{\gamma\gamma}$ & $\Gamma_{\gamma\gamma}(R)$ & \cite{Soni2018} & \cite{Segovia2016} & \cite{Negash2016} & \cite{MShah2012} & \cite{Li2009} & \cite{Ebert2003} & \cite{Godfrey1985} \\
\hline
$1{^1S_0}$ & $0^{-+}$ &  0.3782   & 0.3035    & 0.387    & 0.69   & 0.738  & 0.496   & 0.527   &  0.35   & 0.214 \\

$2{^1S_0}$ & $0^{-+}$  & 0.2636     &  0.2122  & 0.263   & 0.36    & 0.508  & 0.212  & 0.263  & 0.15     & 0.121 \\

$3{^1S_0}$ & $0^{-+}$  &  0.2067   & 0.1668  &  0.229  &  0.27  &  0.426 &   0.135 & 0.172 & 0.10  & 0.906 \\

$4{^1S_0}$ & $0^{-+}$  & 0.1705     & 0.1378 &  0.201  & \ldots  & \ldots & \ldots &   0.099  & 0.105   & 0.755 \\

$5{^1S_0}$ & $0^{-+}$  &  0.1455     & 0.1176 &  0.193  & \ldots & \ldots & \ldots  &  0.078   & 0.121  &   \ldots\\

$6{^1S_0}$ & $0^{-+}$  & 0.1235     & 0.1000 & \ldots   & \ldots & \ldots & \ldots  &     &   &   \ldots\\

$1{^3P_0}$ & $0^{++}$  & 0.0811      & 0.1150      & 0.0196       &  0.12     & \ldots & \ldots & 0.050 & 0.038 & 0.0208\\
$2{^ 3P_0}$ & $0^{++}$  & 0.0717 &  0.1014    & 0.0195       &  0.14    &  \ldots & \ldots& 0.037& 0.029 & 0.0227 \\
$3{^3P_0}$ & $0^{++}$  &  0.0620     & 0.0875      &  0.0194      &  0.15    & \ldots  & \ldots & 0.037 & \ldots & \ldots\\
$4{^3P_0}$ & $0^{++}$  & 0.0545     & 0.0768      &  0.0192      & \ldots     & \ldots  & \ldots& \ldots & \ldots &\ldots\\
$5{^3P_0}$ &$0^{++}$  &  0.0487     & 0.0686     &  \ldots      & \ldots     & \ldots  & \ldots& \ldots & \ldots &\ldots\\

$1{^3P_2}$ & $2^{++}$  &  0.0213    &  0.0147     &   0.0052     & 0.00308     & \ldots  & \ldots& 0.0066 & 0.008 & 0.0051\\
$2{^3P_2}$ & $2^{++}$   &  0.0189     &  0.0131     & 0.0052       & 0.00384     & \ldots  & \ldots& 0.0067 & 0.006 & 0.0062\\
$3{^3P_2}$ & $2^{++}$  & 0.0164     &  0.0114     &  0.0051      &  0.00410    & \ldots  &\ldots & 0.0064& \ldots & \ldots\\
$4{^3P_2}$ & $2^{++}$   &  0.0144     &  0.0100     &  0.0051      &  \ldots    & \ldots  & \ldots & \ldots & \ldots & \ldots\\
$5{^3P_2}$ & $2^{++}$   & 0.0129     &  0.0090     &  \ldots    &  \ldots    & \ldots  & \ldots & \ldots & \ldots & \ldots\\
\hline\hline
\end{tabular}
\end{table*}

\begin{table*}

\tabcolsep 10pt
 \small
\caption{The di-gluon decay widhts (in MeV) of the $S$ wave and $D$ wave bottomonium
 states.} \label{tab7}
\begin{tabular}{ccccccccccccccc}
\hline\hline
% &  &   &   &&&&           & \multicolumn{2}{c} {Experiment} & &&&&\\%
%\cline{9-10}%

\hline
 State  & $J^{PC}$ & $\Gamma_{gg}$ & $\Gamma_{gg}(R)$ & \cite{Soni2018} & \cite{Segovia2016} & \cite{AParmar2010} & \cite{Laverty2009} & \cite{Gupta1996} &  &  \\
\hline
$1{^1S_0}$ & $0^{-+}$ & 5.4496   &  6.8520   &  5.448   &  20.18  & 17.945  & 11.49   &  12.46 &     & \\

$2{^1S_0}$ & $0^{-+}$  & 4.1775     &  5.2374  & 3.710   &  10.64   & \ldots  & 5.16  &  \ldots &      & \\

$3{^1S_0}$ & $0^{-+}$  &  3.4499     &  4.3182 & 3.229   &  7.94  & \ldots  & 3.80  & \ldots  &   &  \\

$4{^1S_0}$ & $0^{-+}$  & 2.9454    &  3.6829 &  2.985  &  \ldots & \ldots  &  \ldots   & \ldots   &   &  \\

$5{^1S_0}$ & $0^{-+}$  &  2.5769   & 3.2196 &  2.832  & \ldots  & \ldots & \ldots    & \ldots  &   &  \\

$6{^1S_0}$ & $0^{-+}$  &   2.2859   & 2.8519 &    & \ldots  & \ldots & \ldots    & \ldots  &   &  \\

$1{^3P_0}$ &  $0^{++}$  & 0.9056    &  1.4297   & 0.276       &  2.00    & 5.250 & 0.96 & 2.15 &\\

$2{^3P_0}$ & $0^{++}$  &  0.7855    & 1.2358      &   0.275     &   2.37   &  \ldots & 0.99 & \ldots& \\
$3{^3P_0}$ & $0^{++}$ & 0.6713     & 1.0539      &   0.273     & 2.46     & \ldots  & \ldots & \ldots & \\
$4{^3P_0}$ & $0^{++}$ &   0.5853    &  0.9175    &  0.271      &  \ldots    & \ldots  & \ldots & \ldots &  \\
$5{^3P_0}$ & $0^{++}$ &  0.5192    &  0.8127   &  \ldots   &  \ldots    & \ldots  & \ldots & \ldots &  \\

$1{^3P_2}$ & $2^{++}$ &  0.2384  &  0.2370  &   0.073     &  0.0836    & 0.822 & 0.33 & 0.22 &\\
$2{^3P_2}$ & $2^{++}$ & 0.2076 & 0.2064 &    0.073    &   0.1042   & \ldots  & 0.35 & \ldots& \\
$3{^3P_2}$ & $2^{++}$  &  0.1778     & 0.1767      & 0.072        & 0.1114     &  \ldots & \ldots& \ldots& \\
$4{^3P_2}$ &  $2^{++}$ & 0.1552    &  0.1543  &   0.072     & \ldots     & \ldots   & \ldots & \ldots & \\
$5{^3P_2}$ & $2^{++}$ &  0.1378    &  0.1370   &   \ldots    & \ldots     & \ldots   & \ldots & \ldots & \\
\hline\hline
\end{tabular}

\end{table*}
\subsection{The Di-gamma decay of bottomonium states}
Adopting the case of the para positronium $(^1S_0)$ state decays to two photons to that of the $^1S_0$ state of quarkonium with the inclusion of color factor, we can get the expression to compute di-gamma decay width of pseudoscalar mesons as \cite{VanRoyen}

\begin{eqnarray}\label{eq:af}
%\resizebox{0.9\hsize}{!}{$%
\Gamma(n^1 S_0\rightarrow \gamma\gamma)=\frac{12 \alpha_e^2 {e_Q^4}}{M_p^2} && |R_{nl}  \left(0\right)|^2 \nonumber \\ && \left[1-\frac{\alpha_s}{ \pi} \left(\frac{20-\pi^2}{3}\right) \right]
%$%
%}%
\end{eqnarray}
where bracketed quantity is the radiative correction \cite{Kwong1988}.

In the case of $P$ wave states, decay width depends on the first derivative of the radial wave function at the origin and is given by \cite{Kwong1988} .

\begin{eqnarray}\label{eq:af}
%\resizebox{1.0\hsize}{!}{$%
\Gamma(n^3 P_0\rightarrow \gamma\gamma)=  \frac{   27   \alpha_e^2 {2^4}}{ {M(n^3 P_0)}^4}{e_Q^4} && |R^{'}_{nl} \left(0\right)|^2 \nonumber \\ &&  \left[1 + \frac{\alpha_s}{\pi} \left(\frac{\pi^3}{3}- \frac{28}{9}\right) \right]
%$%
%}%
\end{eqnarray}

\begin{eqnarray}\label{eq:af}
%\resizebox{0.9\hsize}{!}{$%
\Gamma(n^3 P_2\rightarrow \gamma\gamma)=\frac{36 \alpha_e^2 2^4}{ 5 { {M(n^3 P_2)}^4}}{ e_Q^4} && |R^{'}_{nl} \left(0\right)|^2 \nonumber \\ && \left[1- \left(\frac{16\alpha_s}{3 \pi}\right) \right]
%$%
%}%
\end{eqnarray}
Table \ref{tab6} summarises our computed results for the di-gamma decay width of $n^1S_0$ and $n^3P_{0,2}$ states along with other available data.

\subsection{The Di-gluon decay of bottomonium states}
Similar to the di-photon decay, states holding even charge conjugation can decay into di-gluons. Expression used to compute  $S$ wave spin-singlet state decaying into two gluons together with the radiative QCD correction is given by \cite{Kwong1988}
\begin{eqnarray}\label{eq:af}
%\resizebox{0.9\hsize}{!}{$%
\Gamma(n^1 S_0\rightarrow gg)=\frac{2 \alpha_s^2 }{ 3 m_Q^2} |R_{nl} \left(0\right)|^2  \left[1+\frac{4.4 \alpha_s}{\pi}\right]
%$%
%}%
\end{eqnarray}

conventionally, theoretical expression for di-gamma decay width of the $\chi_{b_{0}}$ and $\chi_{b_{2}}$ states which depends on the first derivative of the radial wave function at origin can be obtained from the potential models \cite{Lansberg2009} and the formula adopted here for the computation with QCD correction factor can be written as \cite{Kwong1988, Barbieri1981}

\begin{eqnarray}\label{eq:af}
%\resizebox{0.65\hsize}{!}{$%
\Gamma(n^3 P_0\rightarrow gg)=\frac{6 \alpha_s^2 2^4}{  { {M(n^3 P_0)}^4}}  |R^{'}_{nl} \left(0\right)|^2  \left[1+\frac{10.0 \alpha_s}{\pi}\right]
%$%
%}%
\end{eqnarray}

\begin{eqnarray}\label{eq:af}
%\resizebox{0.8\hsize}{!}{$%
\Gamma(n^3 P_2\rightarrow gg)=\frac{8 \alpha_s^2 2^4 }{ 5 { {M(n^3 P_2)}^4}}  |R^{'}_{nl} \left(0\right)|^2  \left[1-\frac{0.1\alpha_s}{\pi}\right]
%$%
%}%
\end{eqnarray}

 The calculated di-gluon decay width with the other available theoretical data are listed in Table \ref{tab7}.

\subsection{The Tri-gluon decay of bottomonium states}

The decay width of  $S$ wave vector state annihilates into three gluons along with QCD radiative correction is computed using the relation given by \cite{Kwong1988, Kwong19881}

\begin{eqnarray}\label{eq:af}
%\resizebox{0.9\hsize}{!}{$%
\Gamma(n^3 S_1\rightarrow ggg)=\frac{10 \alpha_s^3 2^2 \left(\pi^2-9\right)}{ 81 \pi { {M(n^3 S_1)}^2}} && |R_{nl} \left(0\right)|^2 \nonumber \\ && \left[1- \frac{4.9 \alpha_s}{\pi}\right]
\end{eqnarray}

And in the case of $n^1 P_1$ state, annihilates to the three gluons is given by \cite{Kwong1988, Kwong19881, Segovia2016}

\begin{eqnarray}\label{eq:af}
%\resizebox{0.9\hsize}{!}{$%
\Gamma(n^1 P_1\rightarrow ggg)=\frac{20 \alpha_s^3 2^4 }{ 9 \pi { {M(n^1 P_1)}^4}} |R^{'}_{nl} \left(0\right)|^2 \ln\left(m_Q \langle r\rangle \right)
\end{eqnarray}
In the case of D-wave, $n^3D_J$ states, the three gluon decay widths are computed using the expression given by \cite{Belanger1987}

\begin{eqnarray}\label{eq:af}
%\resizebox{0.9\hsize}{!}{$%
\Gamma(n^3 D_1\rightarrow ggg)=\frac{760 \alpha_s^3 2^6 }{ 81 \pi { {M(n^3 D_1)}^6}} |R^{''}_{nl} \left(0\right)|^2 \ln\left(4 m_Q \langle r\rangle \right)
\end{eqnarray}

\begin{eqnarray}\label{eq:af}
%\resizebox{0.9\hsize}{!}{$%
\Gamma(n^3 D_3\rightarrow ggg)=\frac{40 \alpha_s^3 2^6 }{ 9 \pi { {M(n^3 D_3)}^6}} |R^{''}_{nl} \left(0\right)|^2 \ln\left(4 m_Q \langle r\rangle \right)
\end{eqnarray}

Table \ref{tab8} contains our predicted three gluon decays for $S$, $P$ and $D$ waves where we have compared our results with other reported values.

\begin{table}

\begin{flushleft}
\tabcolsep 3 pt
 \small
\caption{The Tri-gluon decay widhts (in MeV) of the $S$, $P$ and $D$ wave bottomonium
 states.} \label{tab8}
\begin{tabular}{c c c c c c}
\hline\hline
% &  &   &   &&&&           & \multicolumn{2}{c} {Experiment} & &&&&\\%
%\cline{9-10}%

\hline
$J^{PC}$ & State  & $\Gamma_{ggg}$ & $\Gamma_{ggg}(R)$ & PDG \cite{PDG2018} & \cite{Segovia2016}  \\
\hline
$1{^3S_1}$ & $1^{--}$ &  0.0400   &  0.0285   & 0.0441 & 0.0416   \\

$2{^3S_1}$ & $1^{--}$ &  0.0269   &  0.0193   & 0.0188 & 0.0242   \\

$3{^3S_1}$ & $1^{--}$ &  0.0206   &  0.0148  & 0.0072 & 0.01876   \\

$4{^3S_1}$ & $1^{--}$ &  0.0168   &  0.0121   & \ldots & \ldots   \\

$5{^3S_1}$ & $1^{--}$ &  0.0141   &  0.0102   & \ldots & \ldots   \\

$6{^3S_1}$ & $1^{--}$ &  0.0117   &  0.0085  & \ldots & \ldots   \\

$1{^1P_1}$ & $1^{+-}$ & 0.0357   & \ldots    & \ldots & 0.03526  \\

$2{^1P_1}$ & $1^{+-}$ &  0.0346   &  \ldots  & \ldots & 0.0527   \\

$3{^1P_1}$ & $1^{+-}$ &   0.0331   &  \ldots  & \ldots & 0.0621   \\

$4{^1P_1}$ & $1^{+-}$ &  0.0327   &  \ldots   &  \ldots & \ldots   \\

 $5{^1P_1}$ & $1^{+-}$ & 0.0309   &  \ldots   & \ldots & \ldots   \\

$1{^3D_1}$ & $1^{--}$ &   0.0106   & \ldots    & \ldots & 0.0099 \\

 $2{^3D_1}$ & $1^{--}$ &  0.0119   &  \ldots  & \ldots & 0.0096   \\

$3{^3D_1}$ & $1^{--}$ &   0.0118  &  \ldots  & \ldots & \ldots   \\

 $4{^3D_1}$ & $1^{--}$ &  0.0113   &  \ldots   & \ldots& \ldots   \\

$5{^3D_1}$ & $1^{--}$ &  0.0108   &  \ldots   & \ldots& \ldots   \\

$1{^3D_3}$ &  $3^{--}$ & 0.0060   & \ldots    & \ldots& 0.00022 \\

$2{^3D_3}$ & $3^{--}$ & 0.0056  &  \ldots  & \ldots & 0.00125  \\

$3{^3D_3}$ & $3^{--}$ &  0.0055  &  \ldots  & \ldots & \ldots   \\

$4{^3D_3}$ & $3^{--}$ &  0.0053   &  \ldots   & \ldots &  \ldots   \\

$5{^3D_3}$ & $3^{--}$ &  0.0051   &  \ldots   & \ldots & \ldots   \\

\hline\hline
\end{tabular}
\end{flushleft}
\end{table}

\begin{table}

\begin{flushleft}
\tabcolsep 2.6pt
 \small
\caption{Some other annihilation decay widhts (in keV) of the $S$ and $P$  wave bottomonium
 states.} \label{tab9}
\begin{tabular}{c c c c c}
\hline\hline
% &  &   &   &&&&           & \multicolumn{2}{c} {Experiment} & &&&&\\%
%\cline{9-10}%

\hline
Transition  & $\Gamma$ & $\Gamma(R)$ & PDG \cite{PDG2018} & \cite{Segovia2016}  \\
\hline

$1{^3S_1}\rightarrow \gamma\gamma\gamma $ & $4.6775\times 10^{-4}$  & $ 1.2312\times10^{-4}$  & $ \ldots$ & $3.44 \times10^{-6}$ \\

$2{^3S_1} \rightarrow \gamma\gamma\gamma $  & $3.2849\times 10^{-4}$  & $ 8.9895\times10^{-5}$    &$  \ldots $& $2.00 \times10^{-6}$ \\

$3{^3S_1} \rightarrow \gamma\gamma\gamma  $& $2.5794\times 10^{-4}$  &  $7.2037\times10^{-5}$    &  $\ldots $& $1.55 \times10^{-6}$ \\

$4{^3S_1} \rightarrow \gamma\gamma\gamma  $& $2.1298\times 10^{-4}$  &  $6.0336\times10^{-5}$    &  $\ldots $& $1.29 \times10^{-6}$ \\

$5{^3S_1} \rightarrow \gamma\gamma\gamma  $&  $1.8183\times 10^{-4}$  &  $5.2021\times10^{-5}$   &  $\ldots $& $1.10 \times10^{-6}$ \\

$6{^3S_1} \rightarrow \gamma\gamma\gamma  $& $1.5446\times 10^{-4}$  &  $4.4933\times10^{-5}$   &  $\ldots $& $9.56 \times10^{-7}$ \\

$1{^3S_1} \rightarrow \gamma gg  $& 1.2730  &  0.7220   &  1.18 & 0.79 \\

$2{^3S_1} \rightarrow \gamma gg  $& 0.8689  &  0.4982   &  0.59 & 0.46 \\

$3{^3S_1} \rightarrow \gamma gg  $& 0.6718  &  0.3874   & 0.0097 & 0.36\\

$4{^3S_1} \rightarrow \gamma gg  $& 0.5485  &  0.3176   &  $\ldots$ & 0.30 \\

$5{^3S_1} \rightarrow \gamma gg  $& 0.4646  &  0.2698   & $\ldots$ & 0.25 \\

$6{^3S_1} \rightarrow \gamma gg  $& 0.3894  &  0.2272   &  $\ldots$ & 0.22 \\

$1{^3P_1} \rightarrow q\bar{q} +g  $& 57.9585  &$ \ldots$    & $ \ldots$ & 71.53 \\

$2{^3P_1} \rightarrow q\bar{q} +g  $&  55.3966   &  $\ldots$  &$  \ldots  $& 106.14 \\

$3{^3P_1}  \rightarrow q\bar{q} +g $&  52.9585   &$  \ldots $ & $ \ldots$  & 124.53\\

$4{^3P_1} \rightarrow q\bar{q} +g $&  52.4466   &$  \ldots $  &$ \ldots$ & $\ldots$ \\

 $5{^3P_1}  \rightarrow q\bar{q} +g $& 49.5181   &  $\ldots $  &  $\ldots $& $\ldots $\\

\hline\hline
\end{tabular}
\end{flushleft}
\end{table}
\subsection{Other annihilation channels of vector bottomonium states}

Apart from the decays discussed above, there are other processes by which quarkonium states can annihilate. To elaborate specifically, the decay width of mixed strong and electromagnetic annihilation of $n^3S_1$ states into photon and two gluons is given by \cite{Kwong1987} 

\begin{eqnarray}\label{eq:af}
%\resizebox{0.9\hsize}{!}{$%
\Gamma(n^3 S_1\rightarrow \gamma gg)=\frac{8 {\left(\pi^2-9\right)} \alpha {\alpha_s}^2 {e_Q}^2 2^2} { 9 \pi { {M(n^3 S_1)}^2}} && |R_{nl} \left(0\right)|^2 \nonumber \\ && \left[]1-\frac{7.4\alpha_s}{ \pi}\right]`
\end{eqnarray}
For the spin  triplet state of $S$ wave quarkonium the decay rate into three photon is given by \cite{Kwong1987}
\begin{eqnarray}\label{eq:af}
%\resizebox{0.9\hsize}{!}{$%
\Gamma(n^3 S_1\rightarrow \gamma\gamma\gamma)=\frac{16 {\left(\pi^2-9\right)} \alpha^3 {e_Q}^6 2^2}{ 3  { {M(n^3 S_1)}^2}} && |R_{nl} \left(0\right)|^2 \nonumber \\ && \left[1-\frac{12.6\alpha_s}{ \pi}\right]`
\end{eqnarray}

Also, $n^3P_1$ state decay into light flavour meson and a single gluon. The decay rate for such process can be written as {\cite{Kwong1988}} 

\begin{eqnarray}\label{eq:af}
%\resizebox{0.9\hsize}{!}{$%
\Gamma(n^3 P_1\rightarrow q\bar{q}+g)=\frac{8 \alpha_s^3 n_f 2^4}{ 9 \pi { {M(n^3 P_1)}^4}} |R^{'}_{nl} \left(0\right)|^2 \ln\left(m_Q \langle r\rangle \right)
\end{eqnarray}

Outcomes of these decays are summarised in Table \ref{tab9} along with other available theoretical and experimental data.

\subsection{The Electromagnetic Transition widths of bottomonium states}
Bottomonium states possessed the more compactness in nature due to relatively heavier mass of the bottom quark. In such situation especially dealing with the radiative transition which are governed by an emission or absorption of gamma photon, the wave length of the photon is either larger or comparable to the size of the radiating bottomonium state. So, one expects radiative transition in $b \bar{b}$ dominates. The leading order electromagnetic transitions are electric dipole $(E1)$ and magnetic dipople $(M1)$ transitions.

The selection rules for electric dipole transition $(E1)$ are $\Delta l=\pm1$, $\Delta s=0$. In contrast, for magnetic dipole transitions  $(M1)$, $\Delta l=0$, $\Delta s=\pm1$. Within non relativistic limit, the decay width of the $E1$ transition from the initial state $n_{i} ^{(2s_{i}+1)} {l_{i}}_{J_i}$ to final state $n_{f} ^{(2s_{f}+1)} {l_{f}}_{J_f}$ can be obtained as \cite{Eichten1978}

\begin{equation}
\Gamma (i  \xrightarrow{E1} f+\gamma)=  \frac{4 \alpha  e_Q^2}{3}  (2J^{'}+1) S_{if}^E   \omega^3 \\ |{\cal E}_{if}|^2  \times \frac{E_f}{M_i}
\end{equation}
where
\begin{equation}
\omega =\frac{M_i^2-M_f^2}{2M_i}
\end{equation}
$S^E_{if}$ is the statistical factor
and $\varepsilon_{if}$ is the overlap integral which can be computed using initial and final state wave functions as
\begin{eqnarray}
%\varepsilon_{if}=\frac{3}{\omega}\int_0^\infty dru_{n\ell}(r)u_{n^{'}\ell^{'}}(r)\left[\frac{\omega r}{2}j_0\left(\frac{\omega r}{2}\right)-j_1\left(\frac{\omega r}{2}\right)\right]
{\cal E}_{if} = \frac{3}{\omega}\int_0^\infty && dr u_{n\ell}(r)u_{n^{'}\ell^{'}}(r) \nonumber \\&&\left[{\frac {\omega r} {2}} j_0\left(\frac{\omega r}{2}\right)-j_1\left(\frac{\omega r} {2} \right)\right]
\end{eqnarray}
and
\begin{equation}
S_{if}^E= max (\ell, \ell') \left\{ \begin{array}{ccc}
J&1&J^{'}\\ \ell^{'}&s&\ell\end{array} \right\}^2
\end{equation}

The formula used for the $(M1)$ transition from initial state to final state for the quarkonium system can be given as \cite{Radford2007, Lahde2003}
\begin{eqnarray}\label{eq:M1}
\Gamma(i \xrightarrow{M1} f+\gamma)= && \frac{4 \alpha  {e_Q}^2}{3 {m_Q}^2}  (2J^{'}+1) S_{if}^M \nonumber \\ && \ \omega^3|{\langle f | {j_0} \left( \frac {\omega r} {2}\right)|i\rangle |}^2 \times \frac{E_f}{M_i}
\end{eqnarray}

The statistical factor for the $(M1)$ transition can be given as
\begin{eqnarray}
S_{if}^M= 6 \left(2s +1\right)&&     \left(2s{'}+1 \right) \nonumber \\ && \left\{ \begin{array}{ccc}
J&1&J^{'}\\   s^{'} & l &  s \end{array}   \right\}^2  \left\{ \begin{array}{ccc}
1&\frac{1}{2}& \frac{1}{2}\\ \frac{1}{2}&s^{'}& s\end{array} \right\}^2
\end{eqnarray}
It is to be noted that the term $E_f/M_i$ is acting as the relativistic correction factor to the radiative transition width where $E_f$ is the energy of the final state and $M_i$ being the mass of the initial state. Possible $E1$ and $M1$ transitions including relativistic correction factor are tabulated in Table X and XI along with the available theoretical and experimental data for comparison.
\begin{table*}[htp]
\begin{center}
\tabcolsep 15pt
 \small
\caption{The E1 transition decay widhts (in keV) of bottomonium states.} \label{tab10}
\begin{tabular}{ccccccccccccccc}
\hline\hline
% &  &   &   &&&&           & \multicolumn{2}{c} {Experiment} & &&&&\\%
%\cline{9-10}%

\hline
Initial   & Final   & $\Gamma_{E1}$ & $\Gamma_{E1} \left(R\right)$ & PDG \cite{PDG2018} & \cite{Soni2018} & \cite{Deng2017}& \cite{Segovia2016}& \cite{Godfrey2015} \\
\hline

$1{^3P_2}$ & $1{^3S_1}$ & 15.7013 & 15.0471&\ldots & 57.530 & 31.8 & 39.15   & 32.8  \\
$1{^3P_1}$ & $1{^3S_1}$ & 14.6574 & 14.0602 &\ldots & 54.927  & 31.9  & 35.66   & 29.5  \\
$1{^3P_0}$ & $1{^3S_1}$ & 12.5171 & 12.0327 &\ldots & 49.530 & 27.5 & 28.07   & 23.8 \\
$1{^1P_1}$ & $1{^1S_0}$ & 19.4593 & 18.5864 &\ldots & 72.094 & 35.8 & 43.66   & 35.7 \\
$1{^3D_1}$ & $1{^3P_0}$ & 4.4757 & 4.3472 &\ldots & 9.670 & 19.8  & 20.98   & 16.5  \\
$1{^3D_1}$ & $1{^3P_1}$ & 2.6836 & 2.6123 &\ldots & 6.313 & 13.3  & 12.29   & 9.7 \\
$1{^3D_1}$ & $1{^3P_2}$ & 0.1606  & 0.1564 &\ldots & 0.394 & 1.02  & 0.65   & 0.56 \\
$1{^3D_2}$ & $1{^3P_1}$ & 5.2421  & 5.0988 &\ldots & 11.489 & 21.8  & 21.95   & 19.2 \\
$1{^3D_2}$ & $1{^3P_2}$ & 1.5736 & 1.5321 &\ldots & 3.583 & 7.23  & 6.23   & 5.6 \\
$1{^3D_3}$ & $1{^3P_2}$ & 6.6766 & 6.4976 &\ldots & 14.013 & 32.1  & 24.74   & 24.3 \\
$1{^1D_2}$ & $1{^1P_1}$ & 6.8415 & 6.6567 &\ldots & 14.821 & 30.3  & 17.23   & 24.9 \\
$2{^3S_1}$ & $1{^3P_0}$ & 0.3021 & 0.2968 &1.22 $\pm$ 0.11 & 2.377 & 1.09  & 1.09   & 0.91 \\
$2{^3S_1}$ & $1{^3P_1}$ & 0.6140 & 0.6045 &2.21 $\pm$ 0.19 & 5.689 & 2.17  & 1.84   & 1.63 \\
$2{^3S_1}$ & $1{^3P_2}$ & 0.8412 & 0.8291 & 2.29 $\pm$ 0.20 & 8.486  & 2.62  & 2.08   & 1.88 \\
$2{^1S_0}$ & $1{^1P_1}$ & 0.8877 & 0.8770 &\ldots          & 10.181 & 3.41  &  2.85  & 2.48 \\
$2{^3P_2}$ & $2{^3S_1}$ & 4.5037 & 4.3961 &15.1 $\pm$ 5.6  & 28.848 & 15.3  & 17.50   & 14.3 \\
$2{^3P_1}$ & $2{^3S_1}$ & 4.1508  & 4.0542 &19.4 $\pm$ 5.0 & 26.672 & 15.3  & 15.89   & 13.3 \\
$2{^3P_0}$ & $2{^3S_1}$ & 3.4322 & 3.3572 &\ldots & 23.162 & 14.4  & 12.80   & 10.9 \\
$2{^1P_1}$ & $2{^1S_0}$ & 5.9958  & 5.8377 &\ldots & 35.578 & 16.2  & 17.60   & 14.1 \\
$2{^3P_2}$ & $1{^3S_1}$ & 6.6263  & 6.1221 &9.8$\pm$ 2.3 & 29.635 &  12.5 & 11.38   & 8.4 \\
$2{^3P_1}$ & $1{^3S_1}$ & 6.4644  & 5.9762 &8.9 $\pm$ 2.2 & 28.552 & 10.8  & 9.13   & 5.5 \\
$2{^3P_0}$ & $1{^3S_1}$ & 6.1146  & 5.6606 & \ldots & 26.769 & 5.54  & 5.44   & 2.5 \\
$2{^1P_1}$ & $1{^1S_0}$ & 7.4768  & 6.8844 & \ldots & 34.815 & 16.1  & 14.90   & 13.0 \\
$3{^3S_1}$ & $2{^3P_0}$ & 0.2151  & 0.2128 & 1.20 $\pm$ 0.12 & 3.330  & 1.21  & 1.21   & 1.03 \\
$3{^3S_1}$ & $2{^3P_1}$ & 0.4175  & 0.4137 & 2.56 $\pm$ 0.26 & 7.936 & 2.61  & 2.13    & 1.91 \\
$3{^3S_1}$ & $2{^3P_2}$ & 0.5585  & 0.5536 & 2.66 $\pm$ 0.27 & 11.447 & 3.16   & 2.56   & 2.30 \\
$3{^1S_0}$ & $2{^1P_1}$ & 0.4586  & 0.4555 & \ldots & 13.981 & 4.25  & 2.60   & 2.96 \\
$3{^3S_1}$ & $1{^3P_0}$ & 0.6434  & 0.6122 & 0.055 $\pm$ 0.010 & 0.594 & 0.097  & 0.15   & 0.01 \\
$3{^3S_1}$ & $1{^3P_1}$ &  1.7078 & 1.6277 & 0.018 $\pm$ 0.010 & 1.518 & 0.0005  & 0.16  & 0.05 \\
$3{^3S_1}$ & $1{^3P_2}$ & 2.6850  & 2.5627 & 0.20 $\pm$ 0.03  & 2.354  & 0.14  & .083    & 0.45 \\
$3{^1S_0}$ & $1{^1P_1}$ & 4.3163 & 4.1267 & \ldots & 3.385 &  0.67 & 0.0084   & 1.30 \\
\hline
\hline
\end{tabular}

\end{center}

\end{table*}

\begin{table*}[htp]
\begin{center}
\tabcolsep 15pt
\small
\caption{The M1 transition decay widhts (in eV) of bottomonium states.} \label{tab11}
\begin{tabular}{ccccccccccccccc}
\hline
% &  &   &   &&&&           & \multicolumn{2}{c} {Experiment} & &&&&\\%
%\cline{9-10}%
\hline
Initial   & Final   & $\Gamma_{M1}$ &  $\Gamma_{M1}\left(R\right)$ &   PDG \cite{PDG2018} & \cite{Soni2018} & \cite{Deng2017} & \cite{Segovia2016} & \cite{Godfrey2015} \\
\hline
$1{^3S_1}$ & $1{^1S_0}$ & 3.7986 & 3.7827 & \ldots  & 37.668  &  10  & 9.34    & 10 \\
$1{^3P_2}$ & $1{^1P_1}$ & 0.0197 & 0.0197    & \ldots    & \ldots  & 0.095   &  0.089   & 0.096 \\
$1{^1P_1}$ & $1{^3P_1}$ &  0.0018 & 0.0018  & \ldots  & \ldots  & 0.0094   & 0.0115    & 0.012\\
$1{^1P_1}$ & $1{^3P_0}$ & 0.3191   & 0.3183 & \ldots   & \ldots  & 0.90   &  0.86   & 0.89\\
$2{^3S_1}$ & $2{^1S_0}$ & 1.7856 & 1.7801 &  \ldots  & 5.619  & 0.59   & 0.58    & 0.59\\
$3{^3S_1}$ & $3{^1S_0}$ & 1.0341    & 1.0315 & \ldots   & 2.849  &  3.9  &  0.66   & 0.25\\
\hline
\hline
\end{tabular}

\end{center}

\end{table*}

\section{Results and Discussion}

The spectroscopic masses of S, P and D waves of the bottomonia are computed based on the instanton induced potential in the non-relativistic frame work. The present results are compared with available experimental as well as with other model predictions in Table \ref{tab1} for S wave masses and in Table \ref{tab2} for P wave and D wave masses. Our results are found to be in very good agreement with the experimental values of the respective states. Our estimation for the $1^3S_1$ is 9460.75 MeV which is in  excellent agreement with the PDG listed mass of $1^3S_1 (9460.30 \pm 0.26) $. For the case of spin singlet pseudoscalar state $1^1S_0$ our finding is 9412.22 MeV which is roughly 13 MeV higher than PDG listed mass $(9399.0 \pm 2.3 $ MeV). The mass splitting of 1S state $(1^3S_1-1^1S_0)$ is found to be 48 MeV which is relatively consistent with the PDG listed  mass split for 1S bottomonium (61 MeV). We observe that the S wave mass predictions upto the $4S$ states are very close to the experimental values (PDG average). The deviations observed in the case of 5S and 6S states are somewhat higher with reference to the respective experimental values. Similar agreements are also seen in the case of  P wave states and the D waves with respect to the existing experimental values.  

As far as the identification of the excited bottomonion states are concerned, the masses, decay constant, di gamma, di leptonic decay width are primarily important. Comparing the results of the vector decay constants by our formalism with the experimental result of PDG \cite{PDG2018}  we found that our results are fairly in good agreement. For the orbitally excited states our results with the radiative corrections are in  very good agreement with PDG \cite{PDG2018} reported values. For instance, the vector decay constant of 3S state we have predictrd is $430.42$ MeV and value of vector decay constant reported by PDG \cite{PDG2018} is also $430 \pm 4$ MeV. This indicates that radiative corrections plays an important role in decay mechanism. We found that radiatively corrected results are also compatible with LQCD \cite{Colquhoun2015} predictions. 

In the case of pseudoscalar decay constant, due to the  unavailability of PDG data we have compared our data set with the results of other theoretical model \cite{Soni2018},  Lattice QCD \cite{Davies} and QCD sum rules \cite{Veli2012}. We found our results in good agreement with the Cornell potential predictions \cite{Soni2018}. Also in the case of Lattice QCD, the pseudosalar decay constant for $1^1S_0$ state is  (667 MeV) which is in close agreement with our prediction (654.81 MeV). One can see from the Table \ref{tab4} that predictions from QCD sum rule \cite{Veli2012} is roughly three times lower than all other estimations for $1^1S_0$ state. Also, the results in the case of \cite{Negash2016} (potential model study) are considerably higher than other estimations of the pseudoscalar decay constant. It is important to note that both the vector and the pseudosalar decay constants relie upon  the numerical value of the square of the wave function at the origin, so the choice of the potential and parameters may affect on the predictions.

We have also computed the Di-lepton, Di-gamma  and Di-gluon decay widths and results are summarized in Table \ref{tab5}, \ref{tab6} and \ref{tab7}. Our results are compared with the respective values reported in PDG \cite{PDG2018}. In the case of di-leptonic decay of the $n^3S_1$ it is observed that the radiative corrections are important for higher $(n > 2)$ radial excited states. The results predicted by \cite{Radford2011} (Relativistic potential model) are much closer to our predictions with radiative corrections. Besides that our predicted results with radiative correction are also in a close agreement with \cite{Li2009} (Screened potential model). Our findings are also comparable with other predictions. We have been able to compute the $n^3D_1$ annhilation into electron positron pair but we don't find more estimations inliterature for the detailed comparison.

The di-gamma decay widths are consistent with the outcomes of Cornell potential model \cite{Soni2018} for pseudoscalar $n^3S_0$ states. It is found that the predictions reported by \cite{Negash2016} are almost two times higher than all the other reported values. For the $n^3P_0$ states annihilation into $\gamma \gamma$, our results are comparable with other predictions. Our predictions are found to be higher than other results in the case of $n^3P_2$ states.

For $n^3S_0$ decays into $gg$  our estimated results without radiative correction are consistent with  \cite{Soni2018} and considerably lower than that from the \cite{Segovia2016, Laverty2009, AParmar2010, Gupta1996}. For $n^3P_0$ states, our computed results agree well with the \cite{Laverty2009} without QCD correction while they are in accordance with results of \cite{Segovia2016, Gupta1996} with radiative correction. For these states estimations of \cite{Soni2018} are comparable with our predictions. For the $n^3P_2$ our predictions are in line with that of \cite{Gupta1996} and \cite{Laverty2009} . For the S wave vector state decaying into $ggg$ our results are in agreement with the decay width listed by PDG \cite{PDG2018}. Also they are comparable with the \cite{Segovia2016} especially for $1^3S_1$ state where our prediction is 0.040 MeV without radiative correction and their prediction is 0.041 MeV. For the $n^1P_1$, $n^3D_1$ and $1^3D_3$ we do not include the radiative correction because such correction will be very much lower. The results are consistent \cite{Segovia2016}. Intrestingly, for $1^1P_1$ state our predicted  tri-gluon decay width is 0.035 MeV which is the same as reported by \cite{Segovia2016}. 

Table \ref{tab9} summaries some of the other annihilation decay of the $n^3S_1$ into $\gamma\gamma\gamma$ and $\gamma gg$. As far as the $\gamma\gamma\gamma$  concerns, we find that present results are considerably higher than \cite{Segovia2016}. For  $\gamma gg$ it is again consistent with the PDG \cite{PDG2018} and \cite{Segovia2016}. The radiatively corrected results are allmost matching with the \cite{Segovia2016}.  Results for $ n^3P_1 \rightarrow q {\bar{q}} + g $ are few keV lower than \cite{Segovia2016} but still comparable. For all these decays one has to wait for the experimental confirmation.In general, We can conclude that our instanton potential predictions and those from the constituent quark model predictions \cite{Segovia2016} are in a good agreement for the $ggg$ as well as   $\gamma gg$ decays. 

We present some of the allowed electric dipole transitions (E1) and magnetic dipole transitions (M1) of the first order in Tables \ref{tab10} and \ref{tab11} respectively. Our predicted transition widths are compared with the available other predictions and experimental values wherever it is available. Looking to the results, we find that for $1P \rightarrow 1S $  results are much lower than that reported in \cite{Soni2018}. And similarly we find that our results deviates from \cite{Soni2018} for every transitions except from $1D\rightarrow1P$, $2P\rightarrow1S$ where the predictions do agree. For other transitions like $2P\rightarrow1S$ our predicted outcomes are comparable with the data sets of \cite{Deng2017}, \cite{Segovia2016} and \cite{Godfrey2015}. Particularly, for $2^3P_0$ into $1^3S_1$ our transition width is (5.66 keV) in excellent agreement with the \cite{Deng2017} (5.54 keV) and \cite{Segovia2016} (5.44 keV).  On the experimental side, transition width from $1P\rightarrow1S$, $1D\rightarrow1P$ is not yet listed in PDG. For the magnetic dipole transitions our predictions are more or less comparable with other data set. The discrepancies in the theoretical predictions may vary from one model to the other due to the choice of potential which plays an important role in the predictions of mass which in turn creates differences in phase space which affects the transition widths. Also sometimes it is a choice of different wave function which effects the predictins of the transition widths. Also, instanton vacuum ptential plays a vital role in obtaining mass spectroscopy and other relevent properties of bottomonium. Finally, we hope that our predicted results using instanton effects on heavy quarks will be helpful in search of the new quarkonium physics experimentally as well as theoretically.
\section{Acknowledgement}
We acknowledge the partial support from DST-SERB, India through the major research project: (SERB/F/8749/2015-16).

\end{document}